\documentclass[11pt,onecolumn,draftclsnofoot]{IEEEtran}
\usepackage[latin9]{inputenc}
\usepackage{amsmath}
\usepackage{amsthm}
\usepackage{amssymb}
\usepackage{graphicx}
\usepackage[unicode=true,pdfusetitle,
 bookmarks=true,bookmarksnumbered=false,bookmarksopen=false,
 breaklinks=false,pdfborder={0 0 1},backref=false,colorlinks=false]
 {hyperref}

\makeatletter
\theoremstyle{plain}
\newtheorem{thm}{\protect\theoremname}
\theoremstyle{definition}

\theoremstyle{plain}
\newtheorem{prop}[]{\protect\propositionname}
\theoremstyle{plain}
\newtheorem{lem}[]{\protect\lemmaname}
\theoremstyle{plain}
\newtheorem{cor}[]{\protect\corollaryname}
\theoremstyle{remark}
\newtheorem{rem}[]{\protect\remarkname}

\usepackage{bm}
\usepackage[usenames]{color}

\usepackage{subfigure}

\usepackage{epstopdf}

\usepackage{float}
\usepackage{subfloat}
\usepackage{array}
\usepackage{tabularx}
\usepackage{multirow}
\usepackage{tikz}
\usepackage{algorithmic}

\tikzstyle{arw}=[->,>=latex]
\tikzstyle{node}=[draw,rectangle,rounded corners, minimum width=1cm,minimum height =.75 cm]

\usepackage{adjustbox}

\usepackage{algorithmic}
\usepackage{color}   

\usepackage{hyperref}

\hypersetup{
colorlinks = true,
            linkcolor = blue,
            urlcolor  = blue,
            citecolor = red,
            anchorcolor = blue
}

\providecommand{\corollaryname}{Corollary}
\providecommand{\lemmaname}{Lemma}
\providecommand{\propositionname}{Proposition}
\providecommand{\remarkname}{Remark}
\providecommand{\theoremname}{Theorem}

\makeatother

\providecommand{\corollaryname}{Corollary}
\providecommand{\lemmaname}{Lemma}
\providecommand{\propositionname}{Proposition}
\providecommand{\remarkname}{Remark}
\providecommand{\theoremname}{Theorem}

\begin{document}

\title{Comments on ``Approximate Characterizations for the Gaussian Source
Broadcast Distortion Region''}

\author{Lei Yu, Houqiang Li, \textit{Senior} \textit{Member, IEEE, }and Weiping
Li, \textit{Fellow, IEEE}\thanks{The authors are all with the Department of Electronic Engineering
and Information Science, University of Science and Technology of China,
Hefei, China (e-mail: \{yulei,lihq,wpli\}@ustc.edu.cn). }}
\maketitle
\begin{abstract}
Recently, Tian \emph{et al.} \cite{tian2011approximate} considered
joint source-channel coding of transmitting a Gaussian source over
$K$-user Gaussian broadcast channel, and derived an outer bound on
the admissible distortion region. In \cite{tian2011approximate},
they stated ``due to its nonlinear form, it appears difficult to
determine whether it is always looser than the trivial outer bound
in all distortion regimes with bandwidth compression''. However,
in this correspondence we solve this problem and prove that for the
bandwidth expansion case ($K\geq2$), this outer bound is strictly
tighter than the trivial outer bound with each user being optimal
in the point-to-point setting; while for the bandwidth compression
or bandwidth match case, this outer bound actually degenerates to
the trivial outer bound. Therefore, our results imply that on one
hand, the outer bound given in \cite{tian2011approximate} is nontrivial
only for Gaussian broadcast communication ($K\geq2$) with bandwidth
expansion; on the other hand, unfortunately, no nontrivial outer bound
exists so far for Gaussian broadcast communication ($K\geq2$) with
bandwidth compression.\end{abstract}


\section{Introduction and Preliminaries}

Recently, Tian \emph{et al.} \cite{tian2011approximate} considered
joint source-channel coding (JSCC) of transmitting a Gaussian source
over $K$-user Gaussian broadcast channel, and derived an outer bound
on the admissible distortion region. For $K=2$ case, the properties
of the outer bound were thoroughly investigated by Reznic \emph{et
al.} in \cite{reznic2006distortion}, and in this case, Tian \emph{et
al.} \cite{tian2011approximate} and Reznic \emph{et al.} \cite{reznic2006distortion}
clarified that in certain regimes, this outer bound in fact degenerates
for the case of bandwidth compression, and it is looser than the trivial
outer bound with each user being optimal in the point-to-point setting.
However, the nonlinear form of the bound was cited as the main difficulty
preventing a direct determination whether this outer bound is always
looser than the trivial outer bound in all distortion regimes with
bandwidth compression. Although \cite{tian2011approximate} states
``this outer bound always holds whether the bandwidth is expanded
or compressed'', in this correspondence we prove that for the bandwidth
expansion case (with $K\geq2$), this outer bound is strictly tighter
than the trivial outer bound; while for the bandwidth compression
or bandwidth match case, this outer bound actually degenerates to
the trivial outer bound. It means that on one hand, for Gaussian broadcast
communication ($K\geq2$) with bandwidth compression, no nontrivial
outer bound exists so far; on the other hand, for Gaussian broadcast
communication ($K\geq2$) with bandwidth expansion, the outer bound
given in \cite{tian2011approximate} is nontrivial.

The correspondence is organized as follows. In Section II, we revisit
the outer bound on the admissible distortion region given in \cite{tian2011approximate},
and then prove it to be trivial for bandwidth compression and bandwidth
match cases, and nontrivial for bandwidth expansion case. In Section
III, we give the concluding remarks.

\subsection*{Preliminaries}

The Minkowski inequality given in the following lemma plays an important
role in proving our results.
\begin{lem}[Minkowski inequality]
\cite{hardy1952inequalities}\label{lem:Minkowski} For real numbers
or infinity $0\leq x_{i},y_{i}\leq+\infty,\;i=1,\cdots,n$, and for
$0<p<1$, it holds that
\begin{equation}
\left(\sum_{i=1}^{n}x_{i}^{p}\right)^{\frac{1}{p}}+\left(\sum_{i=1}^{n}y_{i}^{p}\right)^{\frac{1}{p}}\le\left(\sum_{i=1}^{n}\left(x_{i}+y_{i}\right)^{p}\right)^{\frac{1}{p}}.
\end{equation}
Moreover, for $p>1$, the inequality is reversed. In each case equality
holds if and only if the sequences $\left\{ x_{i}\right\} $ and $\left\{ y_{i}\right\} $
are positively linearly dependent (i.e., $y_{i}=\lambda x_{i},\;i=1,\cdots,n$
for some $\lambda\geq0$ or $x_{i}=0,\;i=1,\cdots,n$), or there exists
some $x_{i}$ or $y_{i}$ equal to $+\infty$.
\end{lem}

\section{Main Results }

Consider the problem of broadcasting a Gaussian source $S$ with unit-variance,
i.e., $N_{S}=1$, over a $K$-user Gaussian broadcast channel $Y_{k}=X+Z_{k},k=1,\cdots,K$
with channel noise variances $N{}_{1}>N{}_{2}>\cdots>N{}_{K}>0$ and
transmitting power $P>0$. The encoder maps a source sample block
of length $m$ into a channel input block of length $n$, and each
decoder maps the corresponding channel output block of length $n$
into a source reconstruction block of length $m$. The bandwidth mismatch
factor $b$ is defined as $b=\frac{n}{m}$. For an $\left(m,n,P,d_{1},d_{2},\cdots,d_{K}\right)$
Gaussian source-channel broadcast code with encoding function $f:\mathbb{R}^{m}\rightarrow\mathbb{R}^{n}$
and decoding function $g_{k}:\mathbb{R}^{n}\rightarrow\mathbb{R}^{m}$,
$k=1,\cdots,K$, such that $\frac{1}{n}\sum_{i=1}^{n}\mathbb{E}\left(X\left(i\right)\right)^{2}\leq P$,
the induced distortions are defined as $d_{k}=\frac{1}{m}\sum_{i=1}^{m}\mathbb{E}\left(S\left(i\right)-\hat{S}_{k}\left(i\right)\right)^{2}$,
where $\hat{S}_{k}^{m}=g_{k}(Y_{k}^{n})$ is the source reconstruction
at receiver $k$. A distortion tuple $\left(D_{1},D_{2},\cdots,D_{K}\right)$
is said to be achievable under power constraint $P$ and bandwidth
mismatch factor $b$, if for any $\epsilon>0$ and sufficiently large
$m$, there exist an integer $n\leq mb$ and a Gaussian source-channel
broadcast code $\left(m,n,P,d_{1},d_{2},\cdots,d_{K}\right)$ such
that $d_{k}\leq D_{k}+\epsilon,k=1,\cdots,K$. On this problem, one
outer bound is derived in \cite{tian2011approximate} and shown as
follows.
\begin{thm}
\label{thm:tian}\cite[Thm. 2]{tian2011approximate} Let $0=\tau_{K}\leq\tau_{K-1}\leq\cdots\leq\tau_{1}\leq+\infty$\footnote{For ease of analysis, different from \cite[Thm. 2]{tian2011approximate},
here we allow $\tau_{k}$'s to be infinity. This makes no difference
to the outer bound, since the inequality \eqref{eq:tian} is non-strict.} be any nonnegative real values or infinity. If distortion tuple $\left(D_{1},D_{2},\cdots,D_{K}\right)$
is achievable, then
\begin{equation}
\sum_{k=1}^{K}\Delta N{}_{k}\left[\frac{\left(1+\tau_{k}\right)\prod_{j=2}^{k}\left(D_{j}+\tau_{j-1}\right)}{\prod_{j=1}^{k}\left(D_{j}+\tau_{j}\right)}\right]^{\frac{1}{b}}\le P+N{}_{1},\label{eq:tian}
\end{equation}
where $\Delta N{}_{k}=N{}_{k}-N{}_{k+1},1\leq k\leq K-1$ and $\Delta N{}_{K}=N{}_{K}$.
\end{thm}
In addition, according to cut-set bound, each receiver cannot achieve
a lower distortion than the optimal one in the point-to-point setting,
i.e.,
\begin{equation}
D_{k}\geq D_{k}^{*},k=1,\cdots,K,\label{eq:-7}
\end{equation}
with
\begin{equation}
D_{k}^{*}=\left(\frac{N{}_{k}}{P+N{}_{k}}\right)^{b},k=1,\cdots,K.\label{eq:-6}
\end{equation}
This point-to-point outer bound is referred to as \emph{trivial outer
bound}.

For any $k,1\leq k\leq K$, let $\tau_{K}=\cdots=\tau_{k}=0$ and
$\tau_{k-1}=\cdots=\tau_{1}=+\infty$, then the inequality \eqref{eq:tian}
reduces to \eqref{eq:-7}. Hence we get the following proposition.
\begin{prop}
\label{prop:comp} For any distortion tuple $\left(D_{1},D_{2},\cdots,D_{K}\right)$,
if it satisfies the inequality \eqref{eq:tian} for any $0=\tau_{K}\leq\tau_{K-1}\leq\cdots\leq\tau_{1}\leq+\infty$,
then $D_{k}\geq D_{k}^{*},k=1,\cdots,K$.
\end{prop}
This proposition implies that the outer bound in Theorem \ref{thm:tian}
is tighter than or as tight as the trivial one. However, under what
conditions it is strictly tighter than the trivial outer bound, is
still unknown. Next we will address this problem. First, we consider
the bandwidth compression case, and show that for this case the trivial
outer bound $\left(D_{1}^{*},D_{2}^{*},\cdots,D_{K}^{*}\right)$ satisfies
the necessity given in Theorem \ref{thm:tian}, i.e., $\left(D_{1}^{*},D_{2}^{*},\cdots,D_{K}^{*}\right)$
belongs to the outer bound region given by Theorem \ref{thm:tian}.
\begin{thm}
\label{thm:opt} Let $0=\tau_{K}\leq\tau_{K-1}\leq\cdots\leq\tau_{1}\leq+\infty$
be any nonnegative real values or infinity. If $0<b<1$, then the
distortion tuple $\left(D_{1}^{*},D_{2}^{*},\cdots,D_{K}^{*}\right)$
satisfies the inequality \eqref{eq:tian}.\end{thm}
\begin{IEEEproof}
We adopt mathematical induction to prove Theorem \ref{thm:opt}.

\textbf{Step 1}: For $K=1$, we have
\begin{align}
 & \sum_{k=1}^{1}\Delta N{}_{k}\left[\frac{\left(1+\tau_{k}\right)\prod_{j=2}^{k}\left(D_{j}^{*}+\tau_{j-1}\right)}{\prod_{j=1}^{k}\left(D_{j}^{*}+\tau_{j}\right)}\right]^{\frac{1}{b}}\nonumber \\
= & \Delta N{}_{1}\left(\frac{1}{D_{1}^{*}}\right)^{{\frac{1}{b}}}\\
= & P+N{}_{1}.
\end{align}

\textbf{Step 2}: For $K=2$, we have
\begin{align}
 & \sum_{k=1}^{2}\Delta N{}_{k}\left[\frac{\left(1+\tau_{k}\right)\prod_{j=2}^{k}\left(D_{j}^{*}+\tau_{j-1}\right)}{\prod_{j=1}^{k}\left(D_{j}^{*}+\tau_{j}\right)}\right]^{\frac{1}{b}}\nonumber \\
= & \Delta N{}_{1}\left(\frac{1+\tau_{1}}{D_{1}^{*}+\tau_{1}}\right)^{{\frac{1}{b}}}+\Delta N{}_{2}\left(\frac{D_{2}^{*}+\tau_{1}}{D_{2}^{*}\left(D_{1}^{*}+\tau_{1}\right)}\right)^{{\frac{1}{b}}}\\
= & \Delta N{}_{1}\left(\frac{1+\tau_{1}}{D_{1}^{*}+\tau_{1}}\right)^{{\frac{1}{b}}}+\left(P+N{}_{2}\right)\left(\frac{D_{2}^{*}+\tau_{1}}{D_{1}^{*}+\tau_{1}}\right)^{{\frac{1}{b}}}\label{eq:-3}\\
= & \left(D_{1}^{*}+\tau_{1}\right)^{-{\frac{1}{b}}}\Bigl[\Delta N{}_{1}\left(1+\tau_{1}\right)^{^{{\frac{1}{b}}}}+\left(P+N{}_{2}\right)\left(D_{2}^{*}+\tau_{1}\right)^{{\frac{1}{b}}}\Bigr]\label{eq:-4}\\
= & \left(D_{1}^{*}+\tau_{1}\right)^{-{\frac{1}{b}}}\biggl\{\left[\left(\Delta N{}_{1}\right)^{b}+\left(\Delta N{}_{1}\right)^{b}\tau_{1}\right]^{{\frac{1}{b}}}+\left[\left(N{}_{2}\right)^{b}+\left(P+N{}_{2}\right)^{b}\tau_{1}\right]^{{\frac{1}{b}}}\biggr\}\\
\leq & \left(D_{1}^{*}+\tau_{1}\right)^{-{\frac{1}{b}}}\Bigl[\left(N{}_{2}+\Delta N{}_{1}\right)^{b}+\left(P+N{}_{2}+\Delta N{}_{1}\right)^{b}\tau_{1}\Bigr]^{\frac{1}{b}}\label{eq:-2}\\
= & \left(D_{1}^{*}+\tau_{1}\right)^{-{\frac{1}{b}}}\left[\left(N{}_{1}\right)^{b}+\left(P+N{}_{1}\right)^{b}\tau_{1}\right]^{{\frac{1}{b}}}\\
= & P+N{}_{1},\label{eq:-5}
\end{align}
where \eqref{eq:-2} follows from the Minkowski inequality (Lemma
\ref{lem:Minkowski}).

\noindent \textbf{Step 3}: Assume Theorem \ref{thm:opt} holds for
$K=K'\geq2$. Then for $K=K'+1$, we have
\begin{align}
 & \sum_{k=1}^{K'+1}\Delta N{}_{k}\left[\frac{\left(1+\tau_{k}\right)\prod_{j=2}^{k}\left(D_{j}^{*}+\tau_{j-1}\right)}{\prod_{j=1}^{k}\left(D_{j}^{*}+\tau_{j}\right)}\right]^{\frac{1}{b}}\nonumber \\
= & \Delta N{}_{1}\left(\frac{1+\tau_{1}}{D_{1}^{*}+\tau_{1}}\right)^{{\frac{1}{b}}}+\sum_{k=2}^{K'+1}\Delta N{}_{k}\left[\frac{\left(1+\tau_{k}\right)\prod_{j=2}^{k}\left(D_{j}^{*}+\tau_{j-1}\right)}{\prod_{j=1}^{k}\left(D_{j}^{*}+\tau_{j}\right)}\right]^{{\frac{1}{b}}}\\
= & \Delta N{}_{1}\left(\frac{1+\tau_{1}}{D_{1}^{*}+\tau_{1}}\right)^{{\frac{1}{b}}}+\left(\frac{D_{2}^{*}+\tau_{1}}{D_{1}^{*}+\tau_{1}}\right)^{\frac{1}{b}}\sum_{k=2}^{K'+1}\Delta N{}_{k}\left[\frac{\left(1+\tau_{k}\right)\prod_{j=3}^{k}\left(D_{j}^{*}+\tau_{j-1}\right)}{\prod_{j=2}^{k}\left(D_{j}^{*}+\tau_{j}\right)}\right]^{{\frac{1}{b}}}\\
\leq & \Delta N{}_{1}\left(\frac{1+\tau_{1}}{D_{1}^{*}+\tau_{1}}\right)^{{\frac{1}{b}}}+\left(\frac{D_{2}^{*}+\tau_{1}}{D_{1}^{*}+\tau_{1}}\right)^{\frac{1}{b}}\left(P+N_{2}\right)\label{eq:}\\
\leq & P+N{}_{1},\label{eq:-1}
\end{align}
where \eqref{eq:} follows from the assumption Theorem \ref{thm:opt}
holds for $K=K'$, and \eqref{eq:-1} follows from the formulas \eqref{eq:-4}-\eqref{eq:-5}.
From \eqref{eq:-1}, Theorem \ref{thm:opt} holds for $K=K'+1$.

By combining Steps 1-3, Theorem \ref{thm:opt} holds for any $K\ge1$.
This completes the proof.
\end{IEEEproof}
\noindent Combine Proposition \ref{prop:comp} and Theorem \ref{thm:opt},
then we get the following corollary.
\begin{cor}
\label{cor:compressiontrivial} Assume $0<b<1$. Then for any distortion
tuple $\left(D_{1},D_{2},\cdots,D_{K}\right)$, it satisfies the inequality
\eqref{eq:tian} for any $0=\tau_{K}\leq\tau_{K-1}\leq\cdots\leq\tau_{1}\leq+\infty$,
if and only if $D_{k}\ge D_{k}^{*},k=1,\cdots,K$. \end{cor}
\begin{rem}
Corollary \ref{cor:compressiontrivial} implies that for the bandwidth
compression case, the outer bound given in Theorem \ref{thm:tian}
degenerates to the trivial one.\end{rem}
\begin{IEEEproof}
Actually, the ``only if'' part directly follows from Proposition
\ref{prop:comp}, hence we only need prove the ``if'' part.

Define a function $g\left(D_{1},\cdots,D_{K}\right)$ as
\begin{align}
 & g\left(D_{1},\cdots,D_{K}\right)\nonumber \\
\triangleq & \sum_{k=1}^{K}\Delta N{}_{k}\left[\frac{\left(1+\tau_{k}\right)\prod_{j=2}^{k}\left(D_{j}+\tau_{j-1}\right)}{\prod_{j=1}^{k}\left(D_{j}+\tau_{j}\right)}\right]^{\frac{1}{b}}\\
= & \sum_{k=1}^{K}\Delta N{}_{k}\left[\left(\frac{1+\tau_{k}}{D_{1}+\tau_{1}}\right)\prod_{j=2}^{k}\left(\frac{D_{j}+\tau_{j-1}}{D_{j}+\tau_{j}}\right)\right]^{\frac{1}{b}}.
\end{align}
Obviously, $g\left(D_{1},\cdots,D_{K}\right)$ is monotonically nonincreasing
in $D_{k}$, i.e., $\frac{\partial g}{\partial D_{k}}\le0$. Therefore,
we have $g\left(D_{1},\cdots,D_{K}\right)\le g\left(D_{1}^{*},\cdots,D_{K}^{*}\right)$.
Combining it with Theorem \ref{thm:opt}, i.e., $g\left(D_{1}^{*},\cdots,D_{K}^{*}\right)\le P+N{}_{1}$,
we have $g\left(D_{1},\cdots,D_{K}\right)\le P+N{}_{1}$. It implies
the ``if'' part holds.
\end{IEEEproof}
We can also prove the corresponding results for the bandwidth expansion
case and bandwidth match case.
\begin{thm}
\label{thm:expansionnontrivial} Let $0=\tau_{K}\leq\tau_{K-1}\leq\cdots\leq\tau_{1}\leq+\infty$
be any nonnegative real values or infinity. If $b>1$, then the distortion
tuple $\left(D_{1}^{*},D_{2}^{*},\cdots,D_{K}^{*}\right)$ satisfies
\begin{equation}
\sum_{k=1}^{K}\Delta N{}_{k}\left[\frac{\left(1+\tau_{k}\right)\prod_{j=2}^{k}\left(D_{j}^{*}+\tau_{j-1}\right)}{\prod_{j=1}^{k}\left(D_{j}^{*}+\tau_{j}\right)}\right]^{\frac{1}{b}}\geq P+N{}_{1}.\label{eq:expan}
\end{equation}
Moreover, if $b>1,K\geq2$ and there exists at least one $\tau_{k},1\leq k\leq K$
such that $0<\tau_{k}<+\infty$, then the strict inequality holds
in \eqref{eq:expan}.\end{thm}
\begin{rem}
Theorem \ref{thm:expansionnontrivial} implies that for the broadcast
with bandwidth expansion and at least two receivers, the outer bound
given in Theorem \ref{thm:tian} is (strictly) nontrivial, i.e., it
is strictly tighter than the trivial outer bound.\end{rem}
\begin{thm}
\label{thm:match} Let $0=\tau_{K}\leq\tau_{K-1}\leq\cdots\leq\tau_{1}\leq+\infty$
be any nonnegative real values or infinity. If $b=1$, then the distortion
tuple $\left(D_{1}^{*},D_{2}^{*},\cdots,D_{K}^{*}\right)$ satisfies
\begin{equation}
\sum_{k=1}^{K}\Delta N{}_{k}\left[\frac{\left(1+\tau_{k}\right)\prod_{j=2}^{k}\left(D_{j}^{*}+\tau_{j-1}\right)}{\prod_{j=1}^{k}\left(D_{j}^{*}+\tau_{j}\right)}\right]^{\frac{1}{b}}=P+N{}_{1}.\label{eq:match}
\end{equation}
\end{thm}
\begin{rem}
Theorem \ref{thm:match} implies that for the bandwidth match case,
$\left(D_{1}^{*},D_{2}^{*},\cdots,D_{K}^{*}\right)$ satisfies the
equality \eqref{eq:match} for any $0=\tau_{K}\leq\tau_{K-1}\leq\cdots\leq\tau_{1}\leq+\infty$.
In addition, analog coding could achieve $\left(D_{1}^{*},D_{2}^{*},\cdots,D_{K}^{*}\right)$,
hence for this case, the outer bound given in Theorem \ref{thm:tian}
is tight. Besides, equality \eqref{eq:match} can be also obtained
by examining all inequalities used to derived the outer bound in \cite{tian2011approximate}.
\end{rem}
The proofs of Theorem \ref{thm:expansionnontrivial} and Theorem \ref{thm:match}
are similar to that of Theorem \ref{thm:opt}, and hence omitted here.

\section{Concluding Remarks}

\begin{figure}[t]
\centering\includegraphics[width=0.6\textwidth]{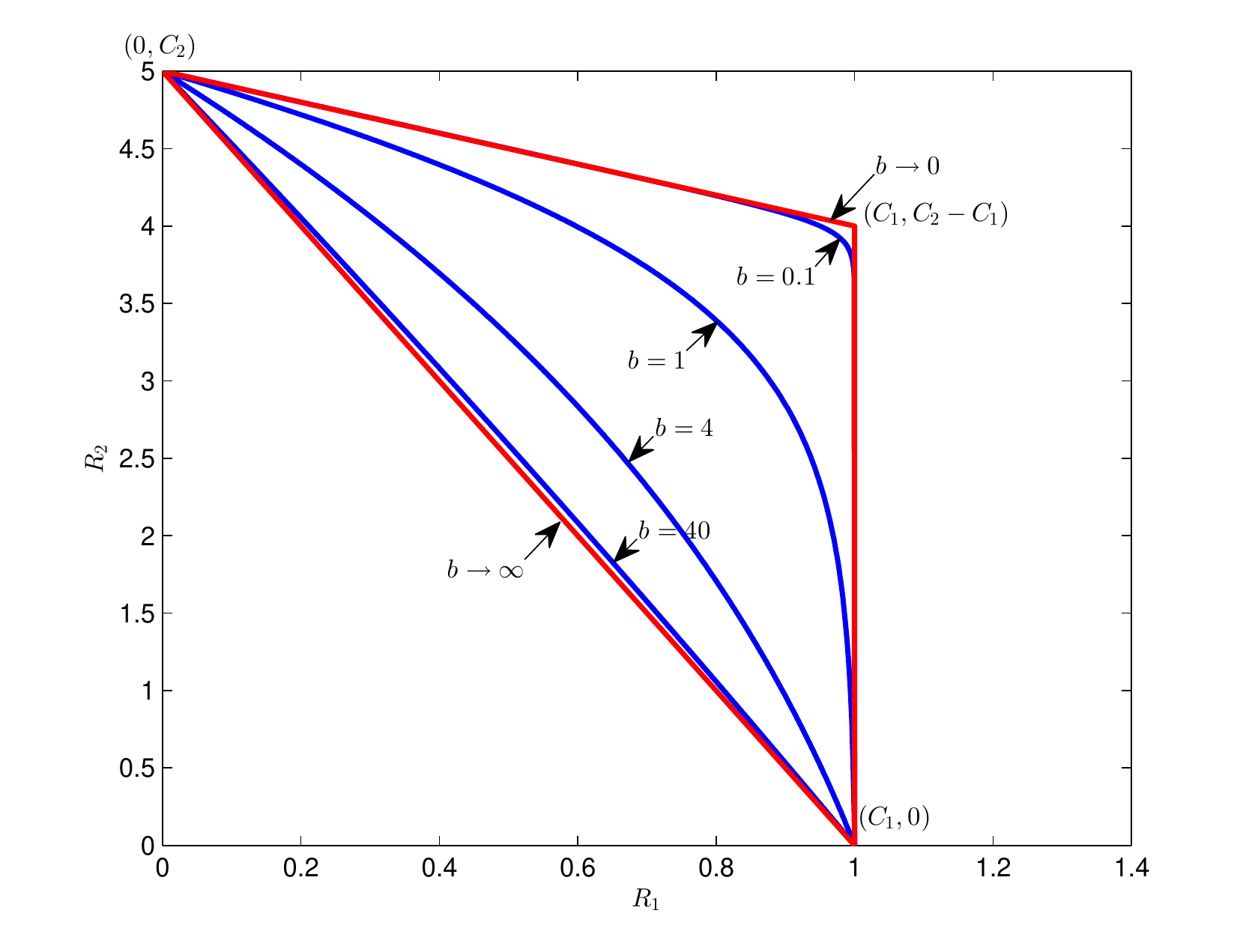} \protect\caption{\label{fig:CapReg}Illustration of the fact that the capacity region
for Gaussian broadcast channel with transmitting power $P$ and noise
variances $N_{1}$ and $N_{2}$ shrinks as the bandwidth $b$ increases
under the point-to-point capacity constraint for each receiver, i.e.,
$\mathcal{C}_{b}(P,N_{1},N_{2})\triangleq\{(R_{1},R_{2}):0\leq R_{1}\leq\frac{b}{2}\log\frac{P+N_{1}}{\alpha P+N_{1}},0\leq R_{2}\leq\frac{b}{2}\log\frac{\alpha P+N_{2}}{N_{1}},0\leq\alpha\leq1\}$
shrinks as $b$ increases under $\frac{b}{2}\log(1+\frac{P}{N_{1}})=C_{1}$
and $\frac{b}{2}\log(1+\frac{P}{N_{2}})=C_{2}$. For this figure,
$C_{1}=1$ and $C_{2}=5$.}
\end{figure}

In this correspondence, we revisited the outer bound on the admissible
distortion region for Gaussian broadcast communication derived in
\cite{tian2011approximate}, and then proved it to be tighter than
the trivial bound for bandwidth expansion case, and degenerate into
the trivial one for bandwidth match and bandwidth compression cases.
It means that on one hand, for Gaussian broadcast communication ($K\geq2$)
with bandwidth compression, no nontrivial outer bound exists so far;
on the other hand, for Gaussian broadcast communication ($K\geq2$)
with bandwidth expansion, the outer bound given in \cite{tian2011approximate}
is nontrivial. Our results lead to a better understanding of this
outer bound, particularly, its relation with the trivial one. In \cite{tian2011approximate},
the outer bound is derived by introducing a set of auxiliary random
variables (or remote sources). The essence of this proof method lies
in that the conditional probability distribution $p_{\hat{S}_{1},\hat{S}_{2},\cdots,\hat{S}_{K}|S}$
can be considered as a virtual broadcast channel (induced by the source
and the reconstructions) realized over the physical broadcast channel
$p_{Y_{1},Y_{2},\cdots,Y_{K}|X}$, hence the capacity region of such
virtual broadcast channel should be contained inside that of the physical
broadcast channel (see \cite{Yu2016,Yu2016-1,Khezeli}). On the other
hand, it can be verified the outer bound given in \cite{tian2011approximate}
is just the necessary condition $\mathcal{C}(N_{S},D_{1},\cdots,D_{K})\subseteq b\mathcal{C}(P,N_{1},\cdots,N_{K})$,
where $\mathcal{C}(N_{S},D_{1},\cdots,D_{K})$ is the capacity region
of the virtual Gaussian broadcast channel with transmitting power
$N_{S}$ and channel noise variances $\frac{N_{S}D_{k}}{N_{S}-D_{k}},1\leq k\leq K$
with $N_{S}=1$ as assumed previously, and $\mathcal{C}(P,N_{1},\cdots,N_{K})$
is the capacity region of the physical Gaussian broadcast channel
with transmitting power $P$ and channel noise variances $N_{k},1\leq k\leq K$.
Therefore, determining whether $\left(D_{1}^{*},D_{2}^{*},\cdots,D_{K}^{*}\right)$
belongs to the outer bound region given in \cite{tian2011approximate}
is equivalent to determining whether $\mathcal{C}(N_{S},D_{1}^{*},\cdots,D_{K}^{*})\subseteq b\mathcal{C}(P,N_{1},\cdots,N_{K})$,
where $\mathcal{C}(N_{S},D_{1}^{*},\cdots,D_{K}^{*})$ is $\mathcal{C}(N_{S},D_{1},\cdots,D_{K})$
with $D_{k}=D_{k}^{*},1\leq k\leq K$. Note that for $D_{k}=D_{k}^{*},1\leq k\leq K$
case, the virtual broadcast channel and the physical broadcast channel
have different bandwidth (the bandwidth ratio is $b$) but the same
point-to-point capacity for each receiver. Combine these with our
results, then we have $\mathcal{C}(N_{S},D_{1}^{*},\cdots,D_{K}^{*})\subseteq b\mathcal{C}(P,N_{1},\cdots,N_{K})$
for $b<1$, $\mathcal{C}(N_{S},D_{1}^{*},\cdots,D_{K}^{*})=b\mathcal{C}(P,N_{1},\cdots,N_{K})$
for $b=1$, and $\mathcal{C}(N_{S},D_{1}^{*},\cdots,D_{K}^{*})\supseteq b\mathcal{C}(P,N_{1},\cdots,N_{K})$
for $b>1$, which further means that the capacity region for Gaussian
broadcast channel shrinks as the bandwidth increases under the point-to-point
capacity constraint for each receiver as shown in Fig. \ref{fig:CapReg}.
This provides an intuitive explanation why the outer bound in \cite{tian2011approximate}
is nontrivial only for bandwidth expansion.

In addition, from the derivation of the outer bound in \cite{tian2011approximate},
it seems that only the one-to-one (continuous) linear analog coding
could achieve this outer bound. However, for bandwidth mismatch case,
the one-to-one continuous mapping does not exist \cite{shannon1949communication,hurewicz1941dimension},
hence we conjecture that for both bandwidth compression case or expansion
case, the outer bound probably cannot be achieved by any source-channel
code.

\bibliographystyle{IEEEtran}

\begin{thebibliography}{1}
\bibitem{tian2011approximate}C. Tian, S. Diggavi, and S. Shamai,
``Approximate characterizations for the Gaussian source broadcast
distortion region,\textquotedblright{} \emph{IEEE Trans. Inf. Theory},
vol. 57, no. 1, pp. 124\textendash 136, 2011.

\bibitem{reznic2006distortion}Z. Reznic, M. Feder, and R. Zamir,
``Distortion bounds for broadcasting with bandwidth expansion,\textquotedblright{}
\emph{IEEE Trans. Inf. Theory}, vol. 52, no. 8, pp. 3778\textendash 3788,
.

\bibitem{hardy1952inequalities}G. H. Hardy, J. E. Littlewood, and
G. P{ó}lya, Inequalities, Cambridge University Press, 1952.

\bibitem{Yu2016}L. Yu, H. Li, and W. Li, ``Distortion bounds for
source broadcast over degraded channel,\textquotedblright{} \emph{IEEE
Int. Symp. Information Theory (ISIT)}, 2016.

\bibitem{Yu2016-1}L. Yu, H. Li, and W. Li, ``Distortion bounds for
source broadcast problem,\textquotedblright{} Submitted to \emph{IEEE
Trans. Inf. Theory}, 2016.

\bibitem{Khezeli} K. Khezeli, and J. Chen ``Outer bounds on the
admissible source region for broadcast channels with correlated sources,''
\emph{IEEE Trans. Inf. Theory}, vol. 61, pp. 4616-4629, Sep. 2015.

\bibitem{shannon1949communication}C. E. Shannon, ``Communication
in the presence of noise,\textquotedblright{} \emph{Proceedings of
the IRE}, vol. 37, no. 1, pp. 10\textendash 21, 1949.

\bibitem{hurewicz1941dimension}W. Hurewicz and H. Wallman, Dimension
theory, Princeton University Press, 1941.\end{thebibliography}

\end{document}